\documentclass[preprint,12pt]{elsarticle}

\usepackage{graphicx}
\usepackage{amssymb}
\usepackage{amsmath}
\usepackage{enumitem}
\usepackage{fancyhdr}
\usepackage{blindtext}

\begin{document}

\journal{Macromolecules}

\begin{frontmatter}


\title{Dewetting of a thin polymer film under shear}



\author[label1]{K. Kadri}
\author[label1]{J. Peixinho}
\author[label2] {T. Salez}
\author[label1]{G. Miquelard-Garnier}
\author[label1]{C. Sollogoub}
\address[label1]{Laboratoire PIMM, Arts et Métiers, CNRS, Cnam, Hesam Université, 151 boulevard de l'Hôpital, Paris, France}
\address[label2]{Univ. Bordeaux, CNRS, LOMA, UMR 5798, F-33405 Talence, France}

\begin{abstract}
The objective of this work is to study the role of shear on the rupture of ultra-thin polymer films. To do so, a finite-difference numerical scheme for the resolution of the thin film equation was set up taking into account capillary and van der Waals (vdW) forces.  This method was validated by comparing the dynamics obtained from an initial harmonic perturbation to established theoretical predictions. With the addition of shear, three regimes have then been evidenced as a function of the shear rate. In the case of low shear rates the rupture is delayed when compared to the no-shear problem, while at higher shear rates it is even suppressed: the perturbed interface goes back to its unperturbed state over time. In between these two limiting regimes, a transient one in which shear and vdW forces balance each other, leading to a non-monotonic temporal evolution of the perturbed interface, has been identified. While a linear analysis is sufficient to describe the rupture time in the absence of shear, the nonlinearities appear to be essential otherwise.
\end{abstract}


\end{frontmatter}


\section{Introduction}
\label{S:1}

Nanolayer coextrusion, an innovative process allowing the combination of at least two polymers in a stratified film or membrane having a total thickness on the order of 100 $\mu$m but composed of thousands of alternating nanometric layers, has gained an increased interest in the past few years \cite{Zhang2019,Zhenpeng2020}. This process offers unique opportunities to explore fundamental questions on the effects of confinement on polymer properties, such as crystallization \cite{Carr2012}, chain mobility and structural relaxation \cite{Arabeche2012, Casalini2016, Monnier2018} or interfacial phenomena \cite{Liu2005a, Liu2005b, Beuguel2019}, as well as to design new nanostructured materials with novel or enhanced properties (mechanical, optical, conductive, gas barrier properties, etc.) \cite{Ponting2010}.
\par However, one strong limitation of the process lies in possible layer breakups, observed by several authors on different polymer pairs when reducing the layer thickness  \cite{Scholtyssek2010,Ho2004,Feng2018}. In a previous study, we investigated this phenomenon in polystyrene (PS)/poly(methyl methacrylate) (PMMA) nanolayered films \cite{Bironeau2017} and evidenced the existence of a critical thickness around 10 nm, below which the layers rupture spontaneously, independently of the processing conditions. We then proposed a mechanism responsible for this layer breakup, similar to the one leading to the dewetting of an ultra-thin polymer monolayer deposited on a solid substrate, as firstly observed by Reiter et al. \cite{Reiter1992} and subsequently explained by Brochard et al. \cite{BrochardWyart1993}.

In nanolayer coextrusion, when the layer thickness is very small - typically below 100 nm-, attractive long-range forces (\textit{i.e.} van der Waals forces) between the two adjacent layers cannot be neglected. Below a critical thickness around 10 nm, they become dominant over the stabilizing capillary forces. In consequence, they may amplify any interfacial instability such as the one due to thermal fluctuations, eventually leading to the layer breakup. Several model experiments on spin-coated three-layer systems were subsequently proposed to confirm this scenario \cite{Zhu2016,Chebil}. Comparing the characteristic dewetting times in a model trilayer system to typical residence times in the nanolayer coextrusion process, we also suggested that the shear induced in the nanolayer process may delay the layer rupture, \textit{i.e.} may stabilize the layers against rupture. 
\par Similar questions have been addressed for many years in the field of fluid mechanics, where the stability of ultra-thin liquid films has been a concern in several industrial applications, such as coating processes \cite{Powell2005} or lithographic printing \cite{Lenz2007}. The stability of thin films has been the subject of many theoretical and experimental studies (see for example the reviews of Oron et al. \cite{Oron1997} 
and Craster et al. \cite{Craster2009}).
The pioneering works of Vrij \cite{Vrij1966} and Sheludko \cite{Sheludko1967} focused on the mechanism of spontaneous rupture of a thin liquid film deposited on a solid substrate. Using slightly different approaches - thermodynamic treatment for Sheludko and diffusion equation for Vrij-, they showed for the first time that the amplification by vdW forces of small irregularities at the film's free surface may lead to a decrease of the total free energy despite the increasing surface, and consequently induce film rupture. They proposed a critical thickness below which the destabilizing vdW forces become dominant over the capillary forces, a critical wavelength of the initial irregularities above which the film is unstable, as well as a growth rate of the perturbation and a characteristic time for rupture. 
\par A more systematic and rigorous approach, developed by Ruckenstein and Jain \cite{Ruckenstein1974}, was based on a linear stability analysis of the Navier-Stokes equations. They assumed that the amplification of small perturbations at the film interface generates a flow in the film. Due to its small thickness, the lubrication approximation was employed. The long-range vdW forces were accounted for through a disjoining pressure term as proposed by Derjaguin \cite{Derjaguin1955}.
Even if this stability analysis is theoretically valid for small perturbations only, information about the conditions leading to film rupture could be obtained. In particular, Ruckenstein and Jain showed that the critical wavelength of the initial periodic disturbance leading to rupture was much larger than the film thickness. These results laid the groundwork for subsequent studies that investigated the nonlinear effects on thin film rupture using either a perturbative analysis \cite{Sharma1986} or numerical computations \cite{Williams1982}.
In the latter study, the authors derived a highly non-linear partial differential equation, a so-called thin film equation, that describes the evolution of the surface of a thin film subject to: i) viscous stresses, ii) a stabilizing Laplace pressure, and iii) a destabilizing disjoining pressure. The main qualitative features of the rupture in these nonlinear studies were similar to the ones in the linear analysis. Still, some quantitative differences were obtained concerning the breakup time in particular, that was found to be systematically inferior in the nonlinear studies compared to the linear analysis. This is likely due to the fact that the latter analysis underestimates the destabilizing effect of the long-range forces. Those various approaches were extended later to multilayer films \cite{Lenz2007,Bandyopadhyay2005,Pototsky2004}.
\par The effect of a shear flow on a thin film rupture was first explored by Kalpathy et al. \cite{Kalpathy2010} for a liquid-liquid interface in a stratified flow and by Davis et al. \cite{Davis2010} in a thin liquid film. They showed that when shear is imposed, the film rupture is delayed and that above a critical shear rate, the rupture is even suppressed. Beyond purely hydrodynamic explanations, another possible effect in practical systems could be the shear-induced modification of the seed thermal fluctuations~\cite{Thiebaud2010,Thiebaud2014}.   \\
In the present study, we investigate the impact of shear on the stability of a polymer thin film, using a numerical approach \cite{Salez2012} inspired by Bertozzi and Zhornitskaya \cite{Bertozzi1998,ZhornitskayaBertozzi2000}. In particular, by  systematically studying various combinations of shear rates and Hamaker constants governing the intensity of vdW forces, we discuss the existence of several regimes for the thin film stability.

\section{Problem position and model}
\subsection{Problem position}
\begin{figure}[!h]
\centering
\includegraphics[width=1.0\linewidth]{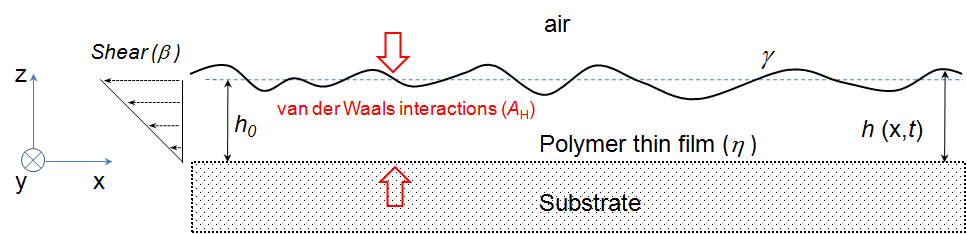}
\caption{A viscous polymer film, of viscosity $\eta$ and nominal thickness $h_0$, is placed atop a rigid substrate. External shear rate $\beta$, surface tension $\gamma$ and van der Waals forces (with Hamaker constant $A_{\textrm{H}}$) compete with each other, and generate a viscous flow that can either result in growth or damping of an initial interfacial perturbation. At horizontal position $x$ and time $t$ the film profile is $h(x,t)$. Invariance along $y$ is assumed.}
\label{fig_schematic_problem}
\end{figure}

The problem studied is represented in Figure \ref{fig_schematic_problem}. A glassy polymer (such as PS) thin film of nominal thickness $h_0$ typically below 100 nm is lying on a substrate and heated well above its glass transition temperature $T_{\textrm{g}}$. For a PS film, $T_{\textrm{g}}$ is about 100$^\circ$C and the temperature of the study, similar to the processing temperature, would be close to 200$^\circ$C. 
At this temperature, the polymer can be considered as a Newtonian fluid with a constant viscosity $\eta_0$, on the order of $10^4$ Pa.s. The viscosity depends on molecular weight, but the value indicated here is typical of polymers used in extrusion \cite{Bironeau2017}.
The surface tension of the polymer with air is noted $\gamma$ ($\sim 27.7$ mN/m for PS at 200$^{\circ}$C \cite{WU1970}) and the Hamaker constant for the substrate/polymer/air system is noted $A_{\textrm{H}}$. 
The value of the latter is difficult to measure experimentally, and though the typical order of magnitude of Hamaker constants is similar for most systems, i.e. $\sim10^{-19}$ J, several values can be found in the literature. For an air/PS/SiO$_2$ system, Seemann et al. \cite{Seemann2001} provide $ A_{\rm H_{air/PS/SiO_2}}= 2.2\,10^{-20}$ J, similar to the value obtained using material refractive indices and dielectric constants from the literature \cite{DeSilva2012,Israelachvili}. 
In the present study, to limit the numerical rupture time which increases with decreasing values of the Hamaker constant, we employ $A_{\rm H}$ to values between 5 10$^{-19}$ and 5 10$^{-18}$ J.
Shear, characterized by a shear rate $\beta$, is applied from right to left. Different values of shear rates will be explored, and their effect on an interfacial perturbation monitored through the evolution of the profile thickness $h(x,t)$ over time $t$ and space along the $x$-axis. 

\subsection{Governing equation}

Taking into account the previous considerations and a spatial invariance along the horizontal direction $y$, the general thin film equation \cite{Oron1997} is assumed to describe the dynamics of the thickness profile $h(x,t) = h_0 + \delta h(x,t)$, where $\delta h(x,t)$ is the perturbation field with respect to $h_0$. The thin viscous film is experiencing Laplace and disjoining pressures, as well as shear stresses. Neglecting gravitational forces, the thin film equation reads in our case:
\begin{equation}
\partial_t h+\frac{\gamma}{3\eta}\partial_x\left(h^3\partial_x^3h + \frac{3{\rm A}_{H}}{6\pi \gamma h}\partial_x h\right) - \beta h\partial_x h=0\ .
\label{eq_tfe_shear}
\end{equation}

Interfacial tension and viscosity are considered constant and possible changes as a function of the film thickness are also neglected in this study.
To solve numerically the equation, one introduces the following dimensionless variables and parameters:
\begin{equation}
\begin{split}
H = \frac{h}{h_0} \quad ; \quad 
\Delta H  = \displaystyle \frac{\delta h}{h_0} \quad
X = \frac{x}{h_0} \quad ; \quad
T  = \frac{\gamma t}{3\eta h_0} \quad ; \quad
\Lambda  = \frac{\lambda}{h_0} \\
K  = kh_0   = \frac{2\pi}{\Lambda} \quad ; \quad
{A} = \frac{{A}_{\rm H}}{6\pi\gamma h_0^2} \quad ; \quad
{B} = \frac{3\eta\beta h_0}{2\gamma}\ ,
\end{split}
\label{Eq2}
\end{equation}
where $\lambda$ is the wavelength of the initial harmonic perturbation (see below) and $k$ is the associated angular wavenumber. Note that both $A$ and $B$ depend on the nominal film thickness and surface tension. The dimensionless thin film equation can then be written as:
\begin{equation}
\partial_T H + \partial_X \left[ H^3\left(\partial_X^3 H +3AH^{-1} \partial_X H\right)\right]-2BH\partial_X H =0\ .
\label{TFE_VDW_adim}
\end{equation}

The parameters of the study and the ranges over which they have been varied are summarized in Table 1.\\

\begin{table}[h!]
\begin{center}
\begin{tabular}{ccc|cccc|cc} 
\hline
\multicolumn{3}{c|}{Parameters}&\multicolumn{2}{c}{Dimensionless parameters}\\
\hline
$h_0$ & $A_{\rm H}$ & $\beta$ & $A$ & $B$\\
nm & J & s$^{-1}$\\
\hline
[10--100] & [5$\times 10^{-19}$--$5\times 10^{-18}$] & [0.2--200] & [0.001--0.1] & [$10^{-5}$--1]\\
\hline
\end{tabular}
\caption{Explored ranges for the parameters of the problem.}
\label{Tableau_parametres_variables_shear}
\end{center}
\end{table}

\subsection{Numerical method and boundary conditions}

The numerical procedure used here is a finite-difference method for thin-film flows \cite{Salez2012}. Specifically, we aim at following the temporal and spatial evolution of an initial harmonic perturbation of wavelength $ \lambda$. In contrast to previous studies (Davis et al. \cite{Davis2010} or Kalpathy et al. \cite{Kalpathy2010}), we do not use periodic lateral boundary conditions here, but a large spatial window size instead. To optimize the computational time and to limit the artificial lateral boundary effects, we consider a truncated initial perturbation, with 1.5 periods, completed at its edges by a flat profile. We impose flat conditions at the boundaries of the numerical domain. Finally, we have checked that the chosen number of periods at the center and the size of the spatial window do not affect the results. 

At $ T = 0 $, the initial profile of the film is spatially discretized over $M$ segments, with a fixed spatial step $\Delta X $ and a spatial index $i\in[0, M-1]$, as:
\begin{equation}
H \left[  \Delta X (i-i_0), 0 \right] = 1 + \Delta H \cos \left[
{2 \pi \Delta X (i-i_ {0})} / {\Lambda} \right]\ ,
\end{equation}
where $i_0$ is the index of the window center.\\ The numerical integration of Equation~(\ref{TFE_VDW_adim}) along time $T$ is then performed using a 4$^{\rm th}$-order Runge-Kutta scheme \cite{Salez2012}. In order to be self-consistent with the lubrication framework and with the window-size constraint above, we impose the following scale separation: $ \Delta X \ll \Lambda \ll M\Delta X $. 

\section{Results}

\subsection{Linear stability analysis}
The effect of shear on the stability of a thin film can be estimated, as a first attempt, using a linear stability analysis. We stress that while such an approach allows one in principle to predict whether an infinitesimal disturbance at the surface is amplified or attenuated, it does not allow for the quantitative study of rupture, which is outside the scope of linearity. 
We consider the evolution of an initial harmonic perturbation of small amplitude $\Delta H_{0}$ around the nominal dimensionless thickness of the film (equals to 1). The perturbative field $\Delta H$ comprises a time-dependent factor $e^{\Gamma T}$, where $\Gamma$ is the growth rate of the perturbation (the perturbation is amplified if $\Gamma$ $>$ $0$ and damped if $\Gamma$ $<$ $0$) and a space-dependent oscillatory factor $e^{iKX}$:
\begin{subequations}
	\begin{align}
		H &=1+\Delta H\\
 		\Delta H&=\Delta H_{0}e^{\Gamma T}e^{iKX}\ .
 	\end{align}
 	\label{deltaH}
\end{subequations}
This leads, after substitution in Equation (\ref{TFE_VDW_adim}), to the following dispersion relation:
\begin{equation}
	\Gamma = K^2(3A-K^2) +2BiK\ .
	\label{DL1 du Gamma}
\end{equation}

\begin{figure}[!h]
\centering
\includegraphics[width=0.6\linewidth]{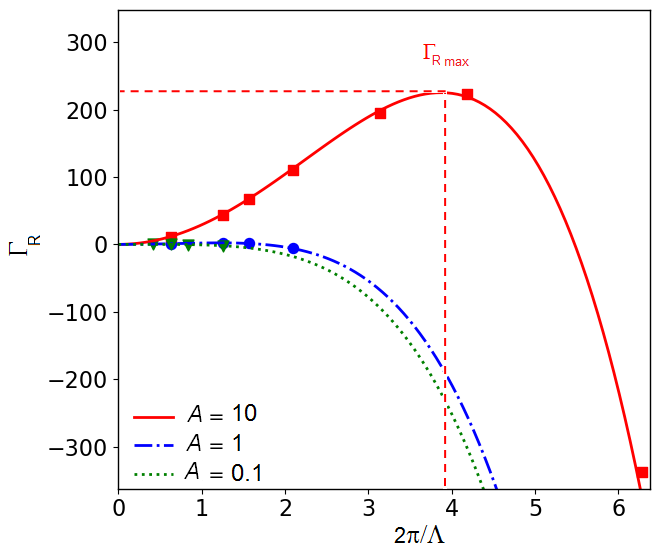}
\caption{Real part $\Gamma_{\textrm{R}}$ of the growth rate of the perturbation as a function of the angular wavenumber $K=2\pi/\Lambda$, as obtained from Equation~(\ref{DL1 du Gamma}) (lines), as well as from  numerical solutions of Equation (\ref{TFE_VDW_adim}) (symbols), for different values of the dimensionless Hamaker constant $A$, as indicated.}
\label{fig_Gamma_K}
\end{figure}

The growth rate of the perturbation is a complex number, with a real part $\Gamma_{\textrm{R}}$ and an imaginary part $\Gamma_{\textrm{I}}$. The real part evaluates the actual rate at which the perturbation is amplified or damped. It appears to be independent of the shear and thus coincides with the solution of the no-shear case, \textit{i.e.} for $B=0$ (see \cite{Ruckenstein1974, Williams1982}). A consequence of such a feature is that a numerical treatment of the non-linear (i.e. beyond linear analysis) thin film equation will be needed in order to understand further the potential role of shear in the dewetting process. The evolution of $\Gamma_{\textrm{R}}$ as a function of $K$, obtained from Equation~(\ref{DL1 du Gamma}), is shown in Figure \ref{fig_Gamma_K} and compared to numerical solutions of Equation (\ref{TFE_VDW_adim}), for different values of $A$, including ones outside the range studied later on with shear. It can be seen that in all cases the numerical results are self-consistently in quantitative agreement with the analytical prediction. Besides, one observes a maximum $\Gamma_{\textrm{max}}=\Gamma(K_{\textrm{max}})$, defined by: 
\begin{equation}
	\begin{split}
		K_{\rm max}=\sqrt{\frac{3A}{2}}\\
		\Gamma_{\rm max} = \frac{9}{4}A^2\ . 	
	\end{split}
	\label{Taux_accroissement}
\end{equation}
Note that the wavelength $\Lambda_{\rm max} = 2 \pi/K_{\rm max} $ of this fastest growing mode will be used as a wavelength $\Lambda$ in all numerical computations below, in order to reduce the total computational time.

Due to the imaginary part $\Gamma_{\textrm{I}}$, the perturbation is transported and shifted in the shear direction. This shift is explicitly highlighted when injecting Equation~(\ref{DL1 du Gamma}) into Equation (\ref{deltaH}), leading to:
\begin{equation}
	\Delta H=\Delta H_{0}\,e^{\Gamma_{\textrm{R}} T}e^{iK(X+2BT)}
	\label{taux_accroissement_reel}\ .
\end{equation} 
Thus, the perturbation propagates along $-X$ (i.e. the shear direction) with a speed equal to $2B$.

\subsection{Rupture without shear}

Here, an initial harmonic perturbation  with an amplitude $\Delta H_0=0.1$ is considered, and Equation (\ref{TFE_VDW_adim}) is solved with $B=0$. Due to vdW forces, the film may undergo a possible rupture after a time which depends on the value of the dimensionless Hamaker constant $A$. Since the numerical scheme is only stable for strictly positive $H$ values, a criterion for rupture has been set as the time $T_{\rm R}$ at which the spatial minimum of $H$ reaches 0.1. We have checked that other small-enough values of this arbitrary threshold do not change qualitatively the results.

\begin{figure}[!h]
\centering
\includegraphics[width=0.85\linewidth]{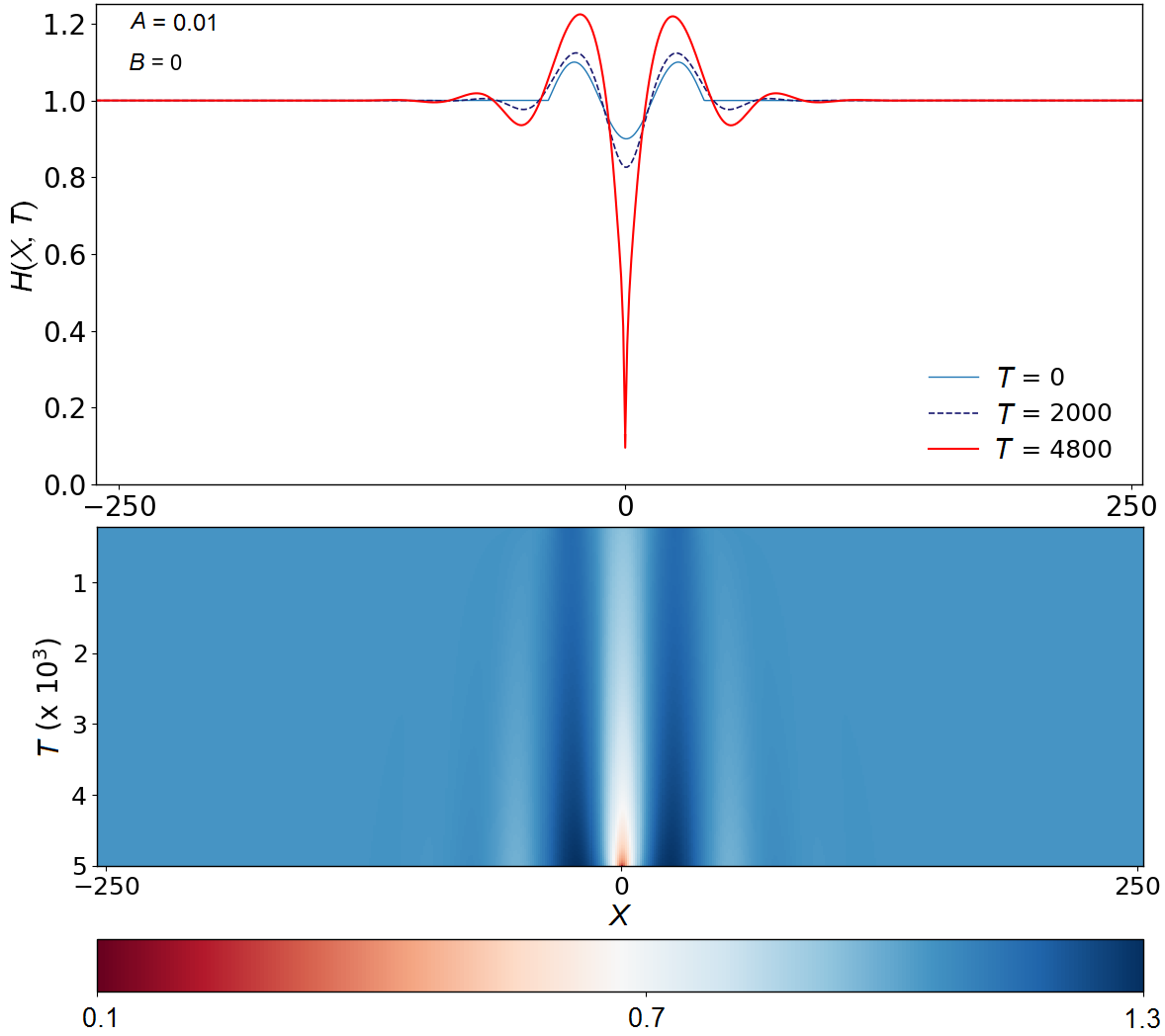}
\caption{(top) Film profiles $H(X,T)$ versus the horizontal position $X$, at three different times $T$ as indicated, as obtained from the numerical solution of Equation (\ref{TFE_VDW_adim}) with $A=0.01$ and $B=0$, for an initial harmonic perturbation with $\Delta H_0=0.1$ and $\Lambda=\Lambda_{\textrm{max}}$. (bottom) Spatiotemporal diagram of the interfacial evolution, with $X$ and $T$ as axes, and the magnitude of $H$ represented using the color code indicated below.}
\label{fig_rupture_no_shear_DST_2D}
\end{figure}

A typical evolution of the film profile is given in Figure \ref{fig_rupture_no_shear_DST_2D} for $A =0.01$. It is seen that the amplitude of the interfacial perturbation increases monotonously over time, until the film ruptures (at $T_{\textrm{R}}=4800$ in this case, with our criterion above). 

\begin{figure}[!h]
\centering
\includegraphics[width=1\linewidth]{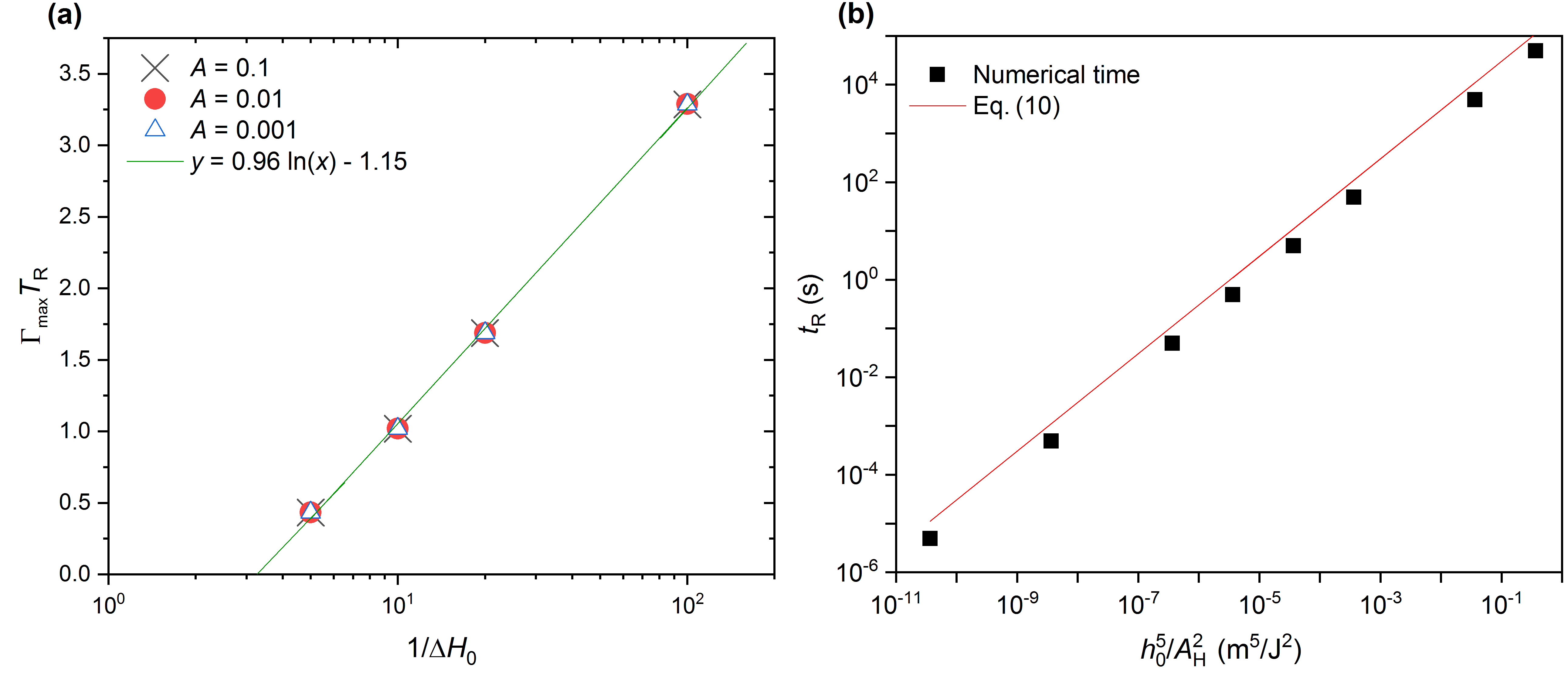}
\caption{(a) Dimensionless rupture time $T_{\textrm{R}}$ as a function of the inverse of the dimensionless perturbation amplitude $\Delta H_0 $, for different values of the dimensionless Hamaker constant $A$. The solid line is a best fit to Eq.~(\ref{Sharma}). (b) Dimensioned version of the rupture times obtained for $\delta h_{\rm 0}$/$h_{\rm 0}$=0.1 in panel a) with the values of interest presented in Table \ref{Tableau_parametres_variables_shear}. The solid line is Eq.~(\ref{formule_Sharma}).}
\label{fig_Sharma_valid}
\end{figure}

We now turn to the detailed study of the rupture time. First, the effect of the initial amplitude of the perturbation is presented in Figure \ref{fig_Sharma_valid}a for different values of $A$. Apart from a numerical prefactor of order unity, the decrease of the dimensionless rupture time $T_{\rm R}$ with increasing initial amplitude $\Delta H_0$ as measured from our numerical solutions follows the linear-stability extrapolation law proposed by Sharma \cite{Sharma1986}:
\begin{equation}
		T_{\rm R}=\frac{1}{\Gamma_{\rm max}} \ln\left(\frac{1}{\Delta H_{\rm 0}}\right)\ .
	\label{Sharma}
\end{equation}

We stress that the expression predicted by Sharma is equivalent to the one by Ruckenstein \cite{Ruckenstein1974} provided that $\Delta H_0=1/e$. Indeed, the expression of Ruckenstein corresponds by definition to the time constant of the exponential growth and is therefore independent of $\Delta H_0$. Similarly, Ruckenstein's expression is similar to the one predicted by Vrij \cite{Vrij1966}, the two expressions quantitatively differing by a factor 2 only. These three theoretical estimates of $T_{\rm R}$ come from a similar linear approximation, and differ only by the exact convention chosen. 

Putting back dimensions, Equation (\ref{Sharma}) is equivalent to:
\begin{equation}
	\begin{split}
    t_{\rm R} & = \frac{48 \pi^2\gamma \eta h_0^5}{A_{\rm H}^2}\ln\left(\frac{h_0}{\delta h_{\rm 0}}\right)\ ,
    \end{split}
    \label{formule_Sharma}
\end{equation}
where $\delta h_{\rm 0}$ is the initial amplitude of the  perturbation. \\
As a guide for practical purposes, and using the values of the physical parameters $\eta$ and $\gamma$ provided above, the rupture time $t_{\textrm{R}}$ is plotted in Figure \ref{fig_Sharma_valid}b as a function of the ratio $h_0^5/A_{\rm H}^2$, with real units, for the case where $\delta h_{\rm 0}$/$h_{\rm 0}$=0.1.

Apart from the numerical prefactor of order unity already mentioned above, it appears that Sharma's prediction describes well the data over 10 decades. This suggests that non-linear effects are not essential to understand the main qualitative features of the film rupture process -- under no shear.

\subsection{Effects of shear}

\subsubsection{Influence of the shear rate}

\begin{figure}[!h]
\centering
\includegraphics[width=0.49\linewidth]{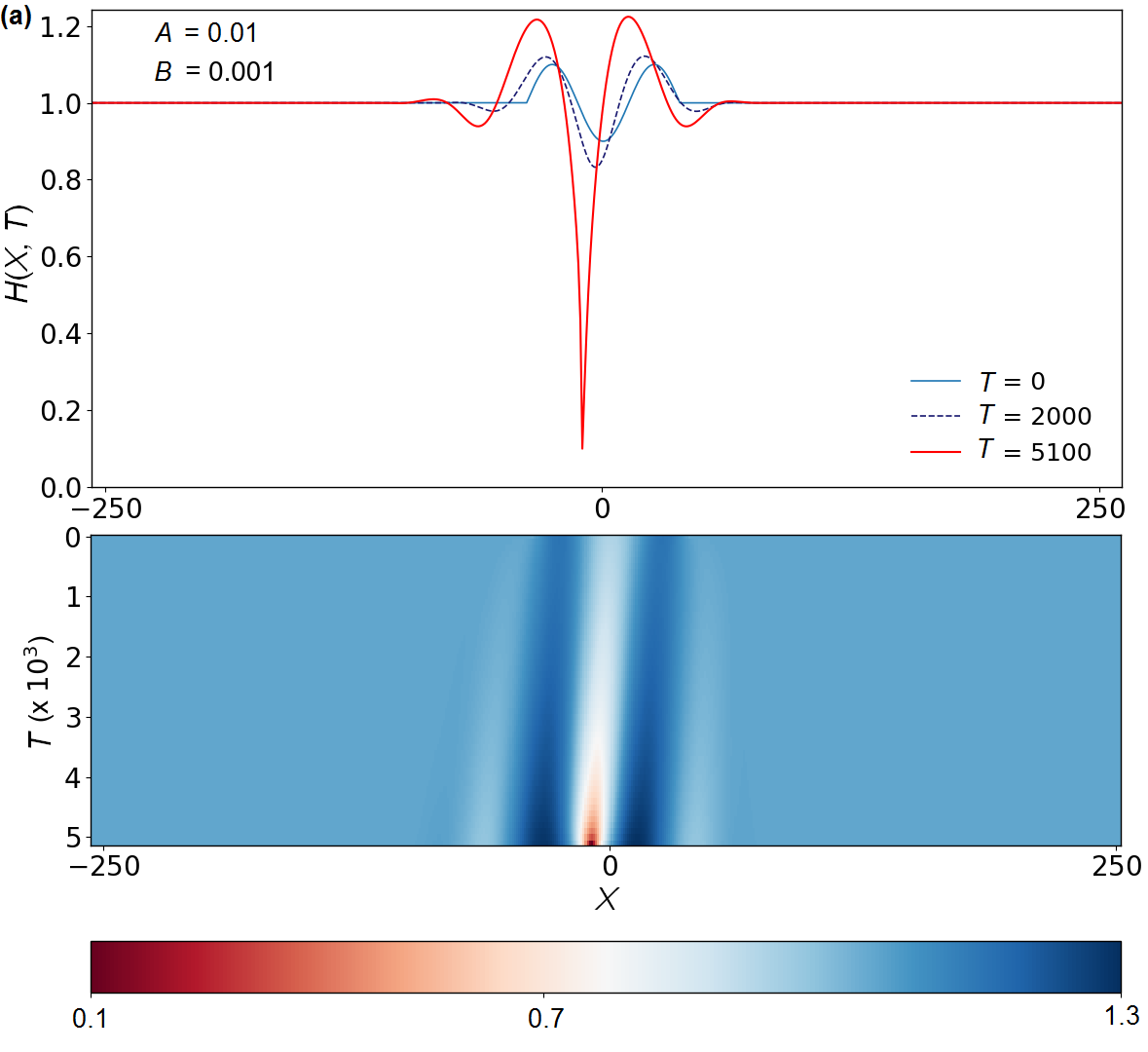}
\includegraphics[width=0.49\linewidth]{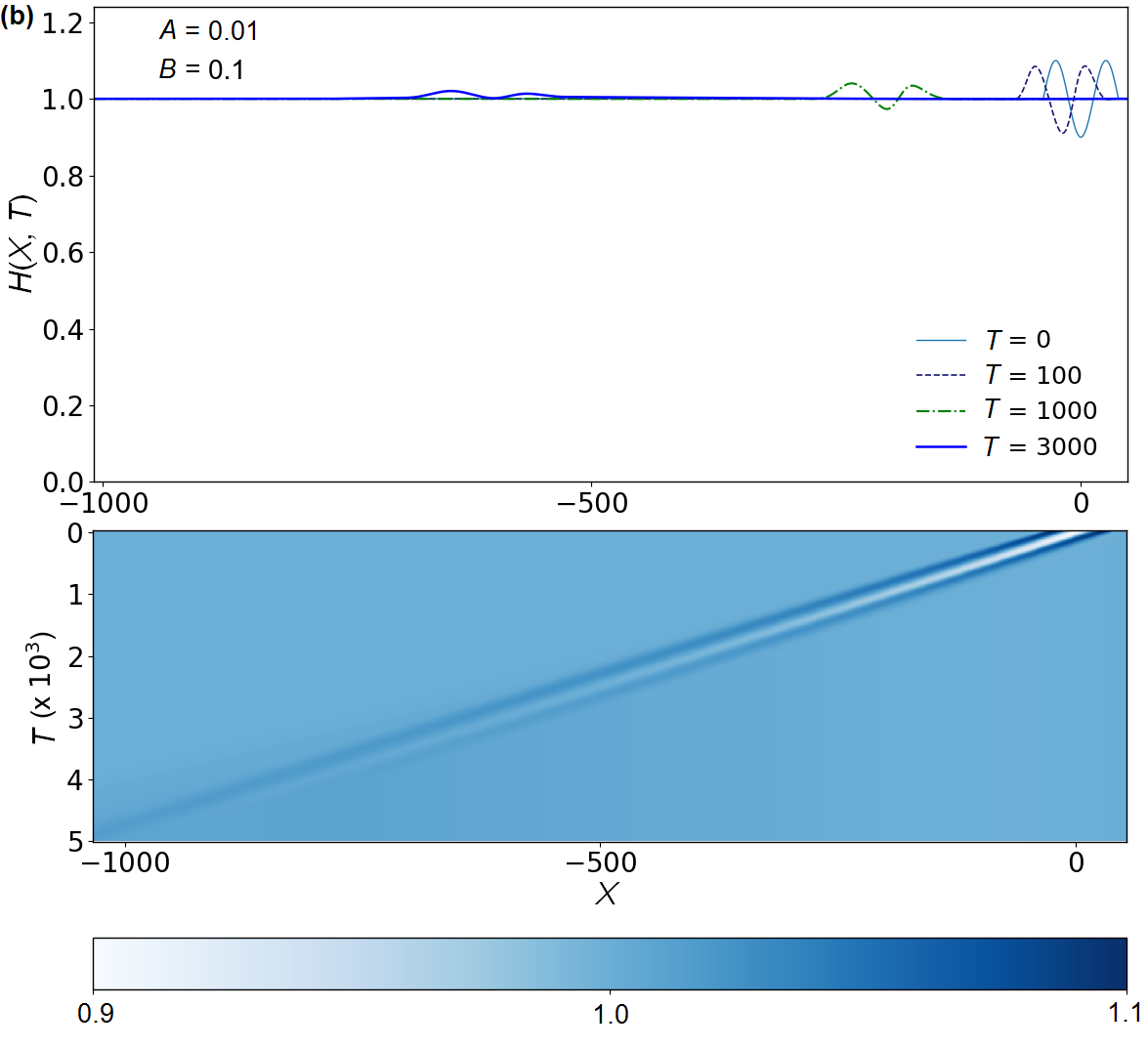}
\caption{(top) Film profiles $H(X,T)$ versus the horizontal position $X$, at three different times $T$ as indicated, as obtained from the numerical solutions of Equation (\ref{TFE_VDW_adim}) with $A=0.01$, for an initial harmonic perturbation with $\Delta H_0=0.1$ and $\Lambda=\Lambda_{\textrm{max}}$. The dimensionless shear rates are fixed to $B=0.001$ and $B=0.1$, in panels (a) and (b) respectively. (bottom) Corresponding spatiotemporal diagrams of the interfacial evolutions, with $X$ and $T$ as axes, and the magnitude of $H$ represented using the color codes indicated below.}
\label{regim of shear}
\end{figure}

Various finite shear rates ($B\neq 0$) have been tested within the same numerical framework as the one described above. The low-shear-rate behavior ($B$ $<<$ $A$) is presented in Figure \ref{regim of shear}a, while the high-shear-rate behavior ($B$ $>>$ $A$) is presented in Figure \ref{regim of shear}b. 
Over time, the perturbation moves along the $x$-axis in the direction in which the shear is applied, from right to left. At low shear rates, the behavior is similar to what is observed without shear: the perturbation grows with time and eventually leads to film rupture. At high shear rates, however, the perturbation is damped, leading to what could be described as a healing of the interface (\textit{i.e.} going back to an unperturbed flat initial state) at long times.

\subsubsection{Critical shear rate}

\begin{figure}[!h]
\centering
\includegraphics[width=0.65\linewidth]{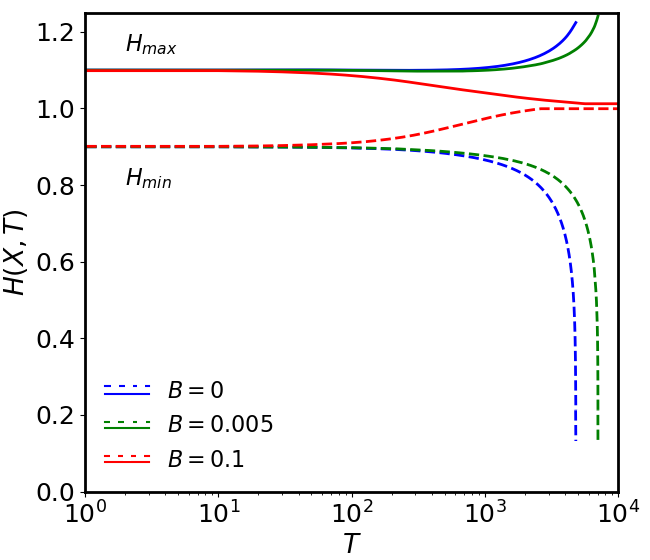}
\caption{Temporal evolution of the  maximum ($H_{\textrm{max}}$) and minimum ($H_{\textrm{min}}$) values of the dimensionless film profile, for three applied dimensionless shear rates, $B = 0, 0.005, 0.1$, as obtained from numerical evolutions such as the ones in Figure~\ref{regim of shear}.}
\label{fig_evolution_hmin_hmax}
\end{figure}

To understand more quantitatively the effect of shear on the perturbed interface profile, the maximum $H_{\textrm{max}}$ and minimum $H_{\textrm{min}}$ of the latter are plotted in Figure \ref{fig_evolution_hmin_hmax} as functions of time, and for different shear rates. At low shear rate, it is seen that the time of rupture increases compared to the case without shear. At high shear rate, healing is confirmed by the fact that both extrema converge to 1, \textit{i.e.} towards the flat-interface situation.\\

\begin{figure}[!h]
\centering
\includegraphics[width=0.85\linewidth]{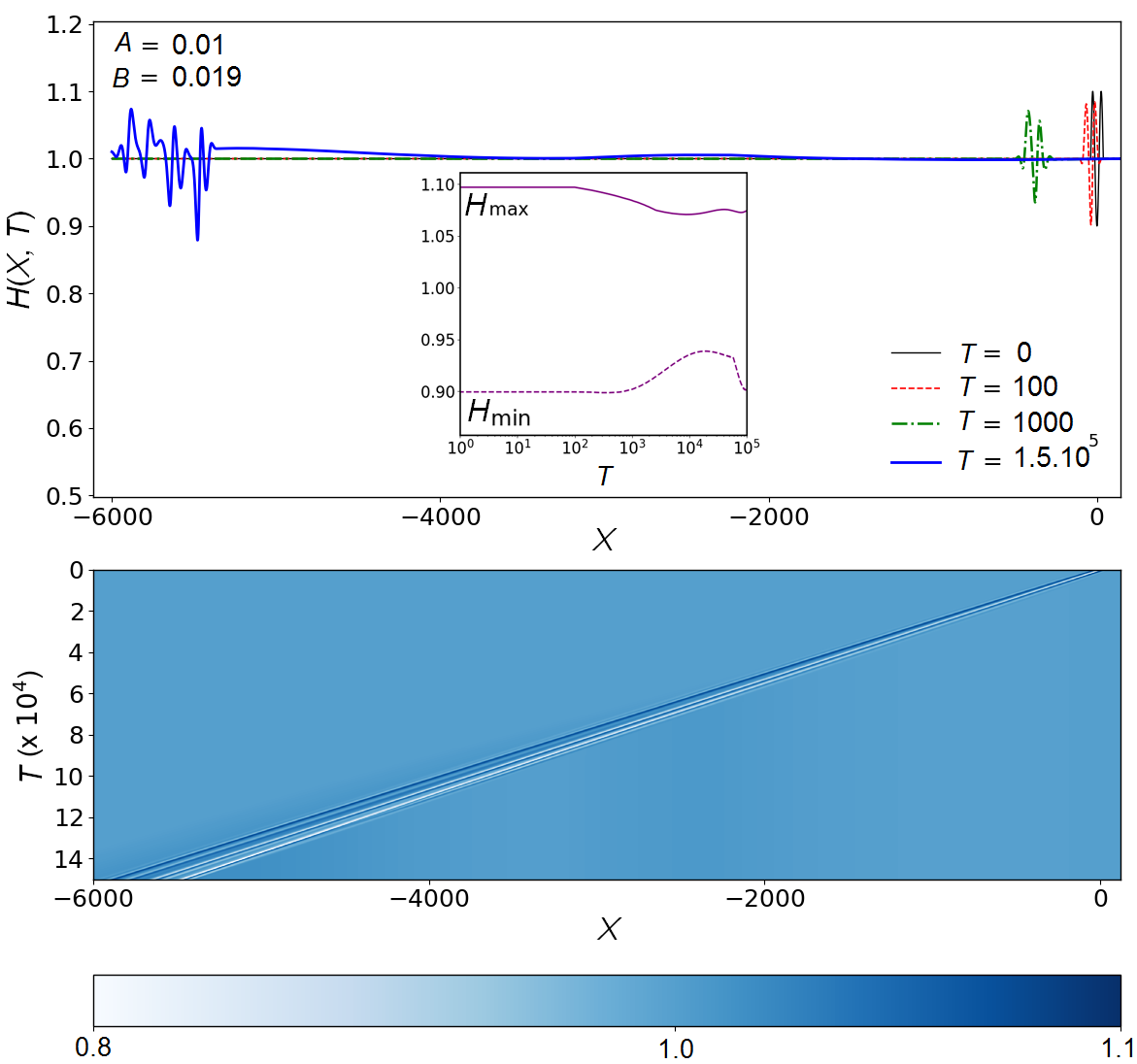}
\caption{(top) Film profiles $H(X,T)$ versus the horizontal position $X$, at four different times $T$ as indicated, as obtained from the numerical solution of Equation (\ref{TFE_VDW_adim}) with $A=0.01$ and $B=0.019$, for an initial harmonic perturbation with $\Delta H_0=0.1$ and $\Lambda=\Lambda_{\textrm{max}}$. Inset: temporal evolutions of the profile's extrema, as in Figure~\ref{fig_evolution_hmin_hmax}. (bottom) Corresponding spatiotemporal diagram of the interfacial evolution, with $X$ and $T$ as axes, and the magnitude of $H$ represented using the color code indicated below.}
\label{fig_transient_regime}
\end{figure}

It is now interesting to examine what happens at intermediate shear rates, \textit{i.e.} $B$ close to $A$. Typical results are presented in Figure \ref{fig_transient_regime}. Here, over the total computational time, no rupture is observed, but no healing either. As seen in the inset, the evolution of the perturbation, characterized as in Figure \ref{fig_evolution_hmin_hmax}, is not monotonic, suggesting the existence of a transient regime where vdW forces and shear compete with each other over times longer than the total time computed.

\begin{figure}[!h]
\centering
\includegraphics[width=0.7\linewidth]{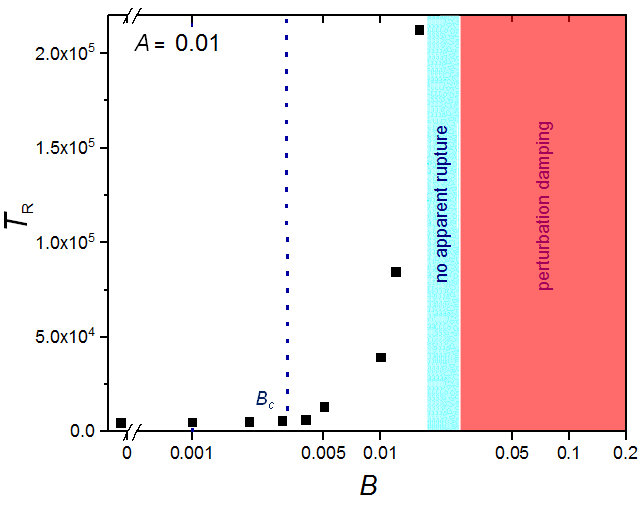}
\caption{Dimensionless rupture time $T_{\rm R}$ as a function of the dimensionless applied shear rate $B$, for a dimensionless Hamaker constant $A = 0.01$. When $T_{\rm R}$ becomes larger than the total computational time, no rupture is observed, and for large enough $B$ healing of the profile is even observed, as summarized by the colored areas. The vertical dashed line indicates $B_{\textrm{c}}\approx3.7\,10^{-3}$, according to Equation~(\ref{bc}).}
\label{fig_evolution_rupture_time}
\end{figure}

We now investigate the influence of shear on the film rupture. Figure \ref{fig_evolution_rupture_time} shows the dimensionless rupture time $T_{\rm R}$ as a function of the dimensionless shear rate $B$. For low shear rates, the rupture time is only slightly higher than the value without shear. Then, for values of $B$ higher than $\sim0.005$, $T_{\rm R}$ increases sharply, and becomes higher than the total computational time for $B>0.014$. For $B>0.03$, perturbation damping and healing of the interface are observed. A so-called "transient regime" is observed for $0.014<B<0.03$. A similar trend was found in a previous numerical simulation (Davis {\it et al.} \cite{Davis2010}) using periodic boundary conditions with a coarse calculation domain. Specifically, Davis {\it et al.} \cite{Davis2010} observed that the rupture is suppressed for $B\approx10 A$. The current systematic study allows us to construct a novel phase diagram, exhibiting in particular: i) perturbation damping for values as low as $B = 3 A$; ii) the existence of a narrow transient regime with a non-monotonic variation of the interface profile along time, resulting in neither rupture nor healing within the accessed temporal and spatial window.\\

Interestingly, a natural dimensionless critical shear rate $B_{\textrm{c}}$ can be identified in the model. Indeed, by balancing the Hamaker and shear contributions in Equation~(\ref{TFE_VDW_adim}), recalling that we have set $K=K_{\textrm{max}}$, and invoking Equation~(\ref{Taux_accroissement}), one gets:
\begin{equation}
    B_{\textrm{c}} \sim 3AK \sim 3\sqrt{\frac32}A^{3/2}\ .
    \label{bc}
\end{equation}
For $A=0.01$, the latter estimate gives $B_{\textrm{c}}\approx0.0037$, which corresponds approximately to the onset value of $B$ in Figure \ref{fig_evolution_rupture_time} after which $T_{\textrm{R}}$ sharply increases with $B$.

\begin{figure}[!h]
\centering
\includegraphics[width=0.8\linewidth]{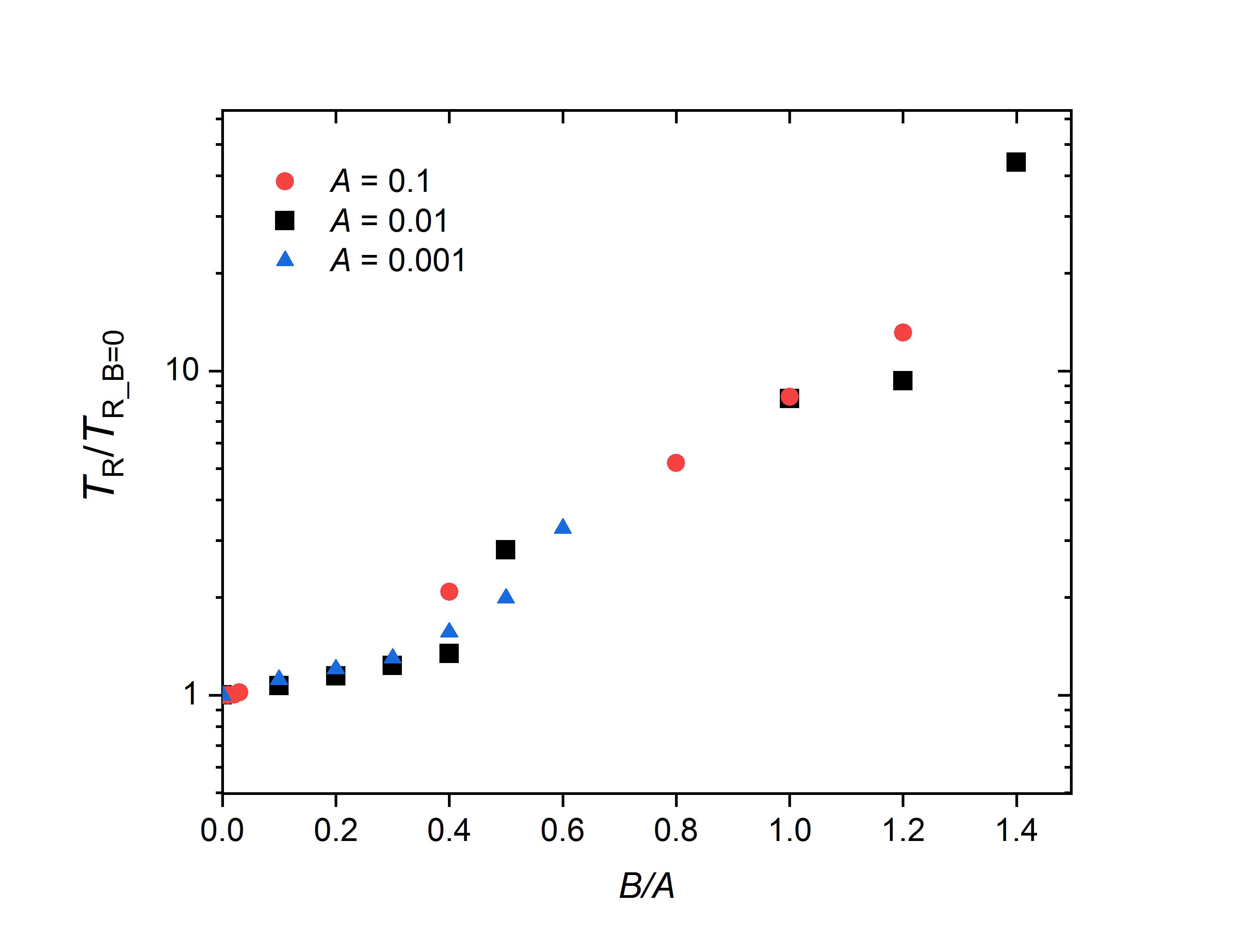}
\caption{Rupture time rescaled with the no-shear rupture time as a function of the ratio between dimensionless shear rate $B$ and dimensionless Hamaker constant $A$, for different values of $A$, as indicated.}
\label{fig_master_curve}
\end{figure}

The study above can be reproduced for several values of $A$. A similar trend is systematically recovered (not shown). Furthermore, a master curve is obtained in Figure \ref{fig_master_curve}, when plotting the ratio of the rupture time with shear and the rupture time without shear as a function of the ratio $B/A$. First, we recover the monotonic increase of the rupture time with shear rate. Secondly, the master rescaling is expected if, near a rupture event,
one neglects the capillary Laplace contribution over the Hamaker one in Equation~(\ref{TFE_VDW_adim}), divide the whole equation by $A$, and absorb $A$ in the definition of time. 

Finally, let us discuss the layer stability in nanolayer coextrusion, from the results obtained here. In our study \cite{Bironeau2017} on a PS/PMMA multilayer system, we made the hypothesis that rupture in multilayer films is induced by thermal fluctuations of amplitude $\sqrt{k_{\rm B}T/\gamma}$ $\sim$ $10^{-9}$ m (where $k_{\rm B}$ is the Boltzmann constant) \cite{Sarlat2006} at the interface between adjacent layers, that are then amplified by vdW forces. Balancing capillary forces with vdW ones, the critical thickness was defined as:
\begin{equation}
    h^* \sim \left(\frac{A_{\rm H}}{3\pi \gamma}\right)^{1/2}\ .
    \end{equation}
Using $A_{\rm H}$ $\sim$ $10^{-18}$ J \cite{DeSilva2012} and and $\gamma$ $\sim$ 1 mN/m \cite{Miquelard2016}, we obtained $h^*$ $\sim$ $10^{-8}$ m, in good agreement with our experimental observations.

In the present study, we showed that rupture can be suppressed for $B\sim3A$. Assuming that our simple model can be employed in the case of a multilayer system too, and putting back dimensions through Equation~(\ref{Eq2}), we get a critical shear rate for rupture suppression:
\begin{equation}
    \beta_{\textrm{s}} \sim \frac{A_{\rm H}}{3\pi \eta h_0^{3}}\ .
    \end{equation}
    
Using $h_0\sim h^*$ and the values of the physical parameters provided above, we obtain $\beta_{\textrm{s}}\approx 10 ~\rm s^{-1}$. Interestingly, the latter value is typical of shear rates occurring during nanolayer coextrusion \cite{Bironeau}. Since $\beta_{\textrm{s}}$ decreases rapidly as $h_0$ increases, it may explain why stable layers with thicknesses as small as 20 nm can be formed via this process -- despite a processing time ($\sim$ 1 min) much larger than the rupture times predicted by Vrij, Ruckenstein and Sharma in a no-shear situation ($\sim$ 1 s, see equation \ref{formule_Sharma}).    

\section{Conclusion}
We have developed a numerical model to study the effect of shear on the stability of an ultra-thin polymer film, taking into account capillary and vdW forces. We identified three regimes: i) a rupture regime at low shear rates, with a rupture time systematically larger than the one in the no-shear case, the latter being in agreement with the expressions predicted by Vrij, Ruckenstein and Sharma; ii) a transient regime in which shear and Hamaker forces compete with each other over the whole time window, leading to a non-monotonic temporal variation of the perturbed interface; iii) a regime at high shear rates in which shear suppresses rupture: a perturbed interface will evolve towards a flat interface over time. Interestingly, while a linear analysis is sufficient to describe the rupture time in the absence of shear, the nonlinearities appear to be crucial in presence of moderate shear. This study paves the way to a better analysis and control of the stabilizing and destabilizing effects in nanocoextrusion processes. 

\section*{Acknowledgements}
The authors would like to thank the Ecole Doctorale Sciences des Métiers de l'Ingénieur (ED SMI 432) for the PhD grant of K.K. and for the 3-month extension during the Covid19 pandemic. The authors also acknowledge funding from the Agence Nationale de la Recherche (grant ANR-21-ERCC-0010-01 EMetBrown) and thank the Soft Matter Collaborative Research Unit, Frontier Research Center for Advanced Material and Life Science, Faculty of Advanced Life Science at Hokkaido University, Sapporo, Japan.

\bibliographystyle{elsarticle-num-names}
\bibliography{references.bib}
\end{document}